\documentclass[fleqn,10pt]{wlscirep}

\usepackage{amsmath}
\usepackage{pifont}
\usepackage{placeins}
\usepackage{booktabs}
\usepackage{tikz}
\usepackage{amssymb}
\usepackage{jabbrv}
\usepackage{graphicx}
\usepackage{jabbrv}
\usepackage{siunitx,etoolbox} 
\robustify\bfseries
\makeatletter
\newcommand{\xleftrightarrow}[2][]{\ext@arrow 3359\leftrightarrowfill@{#1}{#2}}
\makeatother
\newcommand{\xdasharrow}[2][->]{
\tikz[baseline=-\the\dimexpr\fontdimen22\textfont2\relax]{
\node[anchor=south,font=\scriptsize, inner ysep=1.5pt,outer xsep=2.2pt](x){#2};
\draw[shorten <=3.4pt,shorten >=3.4pt,dashed,#1](x.south west)--(x.south east);
}
}

\title{\noindent Mechanics of disordered auxetic metamaterials}
\author[1]{Maryam Hanifpour}
\author[1]{Charlotte F. Petersen}
\author[1]{Mikko J. Alava}
\author[2,3,4,*]{Stefano Zapperi}

\affil[1]{COMP Centre of Excellence, Department of Applied Physics, Aalto University, P.O. Box 11100, FI-00076 Aalto, Espoo, Finland}
\affil[2]{Center for Complexity and Biosystems, Department of Physics, University of Milano, Via Celoria 16, 20133 Milano, Italy}
\affil[3]{CNR - ICMATE, Via R. Cozzi 53, 20125 Milano, Italy}
\affil[4]{CNR - Consiglio Nazionale delle Ricerche, Istituto di Chimica della Materia Condensata e di Tecnologie per l'Energia, 
Via R. Cozzi 53, 20125 Milano, Italy}
\affil[4]{Department of Applied Physics, Aalto University, P.O. Box 11100, FI-00076 Aalto, Espoo, Finland}
\affil[*]{Corresponding author: stefano.zapperi@unimi.it}

\begin{abstract}
Auxetic materials are of great engineering interest not only because of their fascinating negative Poisson's ratio, but also due to the possibility to increase by design the toughness and indentation resistance. The general understanding of auxetic materials
comes mostly from ordered or periodic structures, while auxetic materials used in applications are typically strongly disordered.
Yet, the effect of disorder in auxetics has rarely been investigated. Here, we provide a systematic theoretical and experimental study
of the effect of disorder on the mechanical properties of a paradigmatic two-dimensional auxetic lattice with a re-entrant hexagonal geometry. We show that disorder has a marginal effect on the Poisson's ratio until the point when the lattice topology becomes altered, and in all cases examined the disorder preserves the auxetic characteristics. Depending on the direction of loading applied to these disordered auxetic lattices, either brittle or ductile failure is observed. It is found that brittle failure is associated with a disorder-dependent tensile strength, whereas in ductile failure disorder does not affect strength. Our work thus provides general guidelines to design and optimize elasticity and strength of disordered auxetic metamaterials. 
\end{abstract}

\begin{document}

\flushbottom
\maketitle
%
%
\thispagestyle{empty}


\section*{Introduction}
Auxetic materials display the counter-intuitive property of 
expanding orthogonally to the stretching direction and have some improved mechanical properties, such as toughness and indentation resistance \cite{Article2011,Yang2004}.  This makes them potentially important
for a variety of applications \cite{Evans2000a}, such as bioprostheses \cite{scarpa2008auxetic,caddock1995negative}, acoustic damping \cite{scarpa2004passive,chekkal2010vibro} and aerospace \cite{alderson2007auxetic}. Many of these applications require specific material properties, so precise control and understanding is crucial to realising the full potential of auxetics.
General mechanisms underlying this peculiar behavior have been uncovered in the design of ordered metamaterial lattices \cite{Prawoto2012,Babaee2013,Wu2015,Clausen2015}, 
but their applicability to commercially available bulk auxetic foams, which are typically strongly disordered \cite{Gaspar2005a}, remains unclear.
Previous studies of the effect of disorder yielded a somewhat contradictory picture:
Simple disordered structures designed to imitate bulk auxetics \cite{Lakes1987} include perforated sheets \cite{Grima2016} and arrays of rigid rotating units \cite{Mizzi2015,Pozniak2014,Blumenfeld2011}. In the latter case, disorder has little affect on the Poisson's ratio. In contrast, studies on disordered versions of the re-entrant hexagonal lattice \cite{Bouakba2012,Horrigan2009} demonstrate that disorder can change the impact resistance \cite{Liu2016} and decrease the magnitude of the Poisson's ratio and yield strength significantly \cite{Mukhopadhyay2016,Liu2014}. 

The re-entrant hexagonal geometry is a prototypical ordered auxetic lattice, whose mechanical behaviour has been extensively studied \cite{Gibson1982,Evans1995a,Xu1999,wan2004study,dos2012equivalent}. The auxetic mechanism of this geometry is simple to understand; when tensile load is applied to the lattice, each unit is forced to unfold, thus increasing the samples size on the perpendicular axes. The geometry is auxetic in both orientations, and its properties are sensitive to the geometric cell parameters \cite{Scarpa2000}. In addition to being being a good test case of a simple auxetic mechanism, the structure has potential applications to sandwich panel composites \cite{whitty2003towards,hou2014graded} and textiles \cite{hu2011development}.
Through slight modifications of the structure, control of some of the material properties is possible, such as increasing the in-plane stiffness by adding additional connections \cite{Zied2015}. Further control is gained by using the auxetic honeycomb in combination with other lattices, which allows for control of wave propagation \cite{ruzzene2003control}. Recently, the re-entrant hexagonal lattice has been extended to three dimensions \cite{evans1994auxetic,rad2014analytical} and constructed with 3D printing \cite{Critchley2013,yang2015mechanical,wang2015designable}, direct-laser-writing \cite{buckmann2012tailored} and selective electron beam melting \cite{schwerdtfeger2010auxetic,yang2012non}.

In addition to the elastic properties, the fracture properties of auxetics are of interest because of possibility of an increased fracture toughness relative to conventional materials \cite{Article2011,Donoghue2009}. It has been shown computationally that the ordered re-entrant hexagonal lattice  exhibits improved fracture properties over an equivalent conventional hexagonal lattice \cite{Whitty2003}, however little work exists on the failure of this lattice experimentally \cite{Metamaterials2013}. The presence of disorder has not been considered, even though normally disorder has a key role in fracture strength and toughness \cite{Alava2006}. 

Here we study the elastic properties as well as the fracture mechanics of perfect (zero disorder) and disordered auxetic lattices. We introduce a methodology for adding disorder including the possibility to alter the topology of the lattice, which we find to be crucial for the auxetic behavior. In particular, the Poisson's ratio is shown to depend very strongly on the presence of topological defects. This is remarkable in view of the increasing interest in the role of topology in mechanical metamaterials \cite{Chen2014,Nash2015,Paulose2015}.

\section*{\label{sec:methods}Methods}
\subsection*{\label{sec:constuction}Lattice construction}

\begin{figure}[ht]
\begin{centering}
\includegraphics[width=10cm]{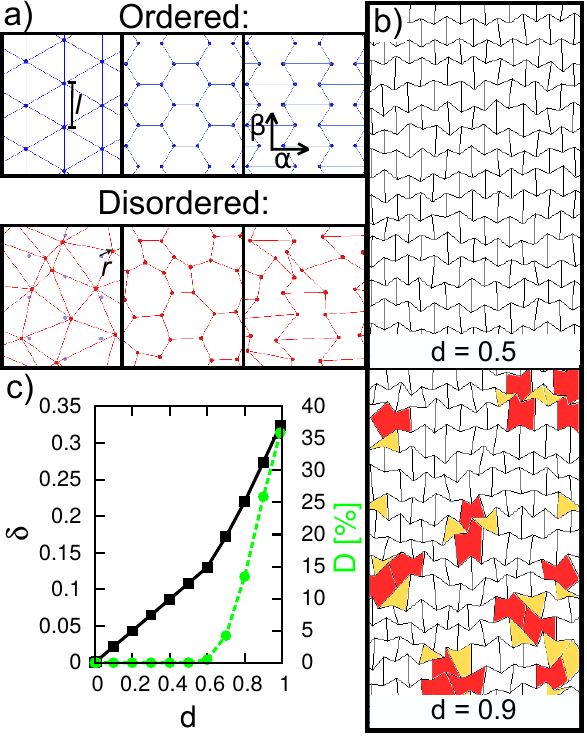}
\par\end{centering}
\caption{a) \textbf{Construction of re-entrant lattices.} The top row shows  a triangular lattice with edge length $l$ (left), its hexagonal dual lattice (middle), and the re-entrant lattice (right). The bottom row shows the disordered triangular lattice (left), where each vertex has been shifted from its ordered position by $\mathbf{r}=[ld_{x}\sqrt{3}/4,ld_{y}/4]$, where $d_{x}$ and $d_{y}$ are uniformly distributed random numbers between -$d$ and $d$. Its dual lattice (middle) is converted to the disordered re-entrant lattice (right) by moving each vertex such that its most horizontal bond is lengthened by 50\%. b) \textbf{Effect of disorder.} Examples of re-entrant lattices with different values of the disorder. Topological defects are colored in red (heptagons) and yellow (pentagons). c) \textbf{Characterisation of disorder.} The normalized standard deviation of the lattice cell area ($\delta$) is plotted as black squares, for various values of the disorder parameter $d$. The percentage of defects in each lattice ($D$) is plotted as green circles.  \label{fig:dis}}
\end{figure}

In previous papers on disordered re-entrant hexagonal lattices \cite{Bouakba2012,Horrigan2009,Liu2016,Mukhopadhyay2016,Liu2014}, no standard method of adding disorder is adopted. Here, we opt for a method of randomly generating lattices which produces geometries that contain disorder, but still maintain characteristics of the re-entrant lattice.  We begin by constructing a disordered triangular lattice, and convert it to a re-entrant hexagonal lattice using the process shown in Fig. \ref{fig:dis}(a). That is, starting from a triangular lattice, we take the dual lattice, which is a hexagonal lattice. Each vertex is then moved such that its most horizontal bond is lengthen by 50\%. If the initial triangular lattice is ordered, this algorithm produces the ordered re-entrant hexagonal lattice. If the initial triangular lattice contains disorder, so does the resultant re-entrant hexagonal lattice. A disordered triangular lattice is generated by shifting the vertex positions of an ordered triangular lattice with edge length $l$ randomly by an amount $\mathbf{r}=[ld_{x}\sqrt{3}/4,ld_{y}/4]$, where $d_{x}$ and $d_{y}$ are uniformly distributed random numbers between -$d$ and $d$. Links between the vertices are then added with a Delaunay triangulation.  The parameter d controls the amount of disorder in the initial triangular lattice, which determines the structure of the disordered re-entrant lattice. This algorithm can produce infinitely many geometries for each level of disorder. All cells in the ordered lattice are hexagons, however variations in the topology are possible in the disordered geometries. Figure \ref{fig:dis}(b) shows examples of disordered lattices with different values of d, and demonstrates that the topology of the lattice is also controlled by this parameter, with only very disordered samples including topological defects. The percentage of topological defects, $D$, is plotted against the disorder parameter $d$ in Fig. \ref{fig:dis}(c). The disorder measure used here, $d$, is specific to the algorithm we have used to generate the disordered geometries. To further characterize the disorder, we compute the area of every cell in the lattice, $A$, and calculate its coefficient of variation $\delta = \left.\sqrt{\left(\left<A^2\right>-\left<A\right>^2\right)}\middle/ \left<A\right>\right.$, where $\left<A\right>$ is the mean of $A$. The parameter $\delta$, plotted in Fig. \ref{fig:dis}(c) as a function of $d$, could be used to quantify disorder in generic two-dimensional random lattices. Fig. \ref{fig:dis}(c) shows that in our case either $d$ or $\delta$ can equivalently be used to quantify disorder.

\clearpage
\subsection*{Experiment}
\begin{figure}[hp]
\begin{centering}
\includegraphics[width=10cm]{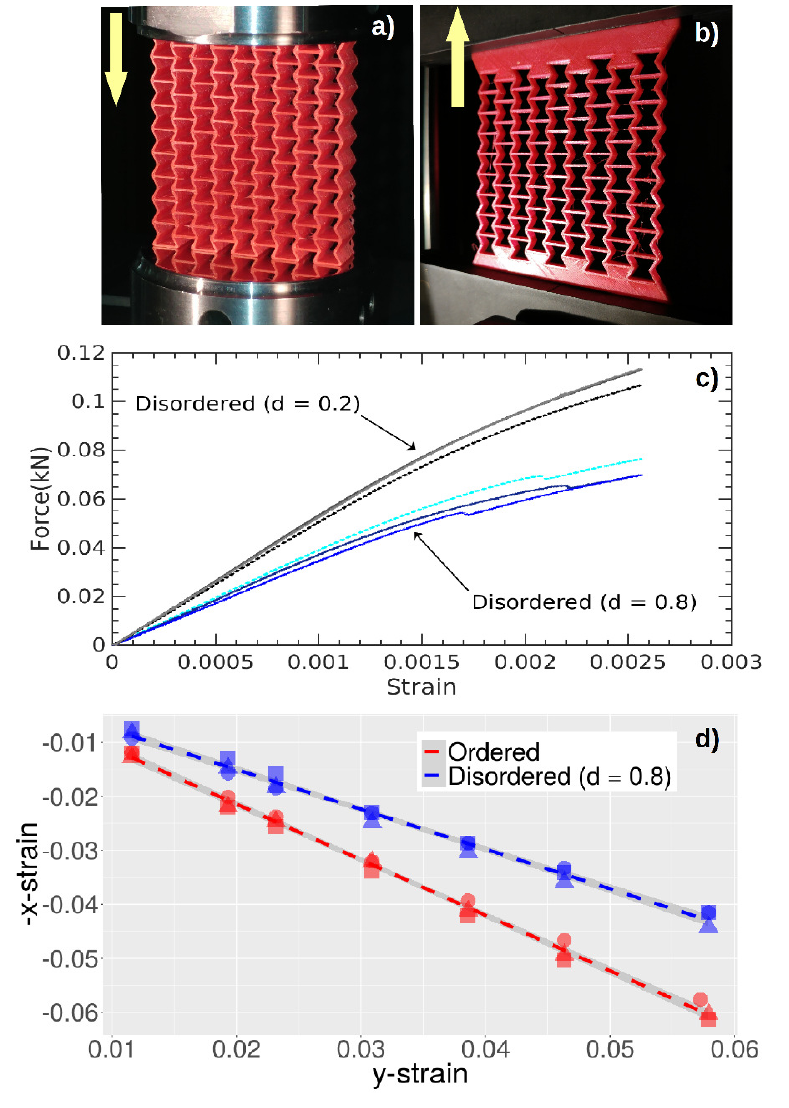}
\par\end{centering}
\caption{ a) \textbf{Experimental tension setup.} b) \textbf{Experimental compression setup.} Force is applied in the direction of the yellow arrow. c) \textbf{Consistency of force measurements.} Force strain plots of three reprinted samples of two example geometries ($d=0.2$ and $d=0.8$). The samples are printed in hard PLA and the force is measured under a tensile load. c) \textbf{Consistency of strain measurements.} -x-strain vs y-strain plots for three reprints of an ordered and a disordered geometry (in the NinjaFlex material). Note that the data points overlap in each case. The load was applied in tension.  \label{fig:exp}}
\end{figure}

Experimentally, samples are prepared for tension [Fig. \ref{fig:exp}(b)] and compression [Fig. \ref{fig:exp}(a)] through 3D printing. The force is applied in one of two orientations, 
labelled as $\alpha$ and $\beta$ in Fig. \ref{fig:dis}(a). 
The 3D models of each geometry have been produced with a bond width of 0.4 mm. Samples are manufactured by means of 3D printing, fused deposition modeling technique (FDM). In this method structures
are produced by laying down many successive thin layers of molten plastic. Each layer thickness is 0.15 mm. The printed samples are of high quality, as evident by the repeatability of mechanical
measurements on reprints of the samples. The force strain curves of reprinted samples of the same geometry are plotted in Fig. \ref{fig:exp}(c), for two different geometries, $d = 0.2$ and $d = 0.8$.
The samples of the same geometry have very similar curves. The strain functions of both a disordered and ordered geometry have the same consistency between samples, plotted in Fig. \ref{fig:exp}(d).
These results demonstrate that the fluctuations due to printing imperfections are minimal, and so variation in physical properties can be attributed to variation in the geometry. 

Three different materials have been used in this work: NinjaFlex (used unless stated otherwise), which is a formulated thermoplastic polyurethane material with super elastic properties; hard PLA; and FlexPLA, which is more flexible compared to hard PLA. The measured mechanical properties of the base material are presented in Table \ref{tab:base}. Due to the force limitations of the testing machine, the Poisson's ratio under compression could not be measured in the solid samples of PLA. More details on material properties at http://vexmatech.com/. Hard PLA is brittle at room temperature, which allows us to study the fracture of the auxetic structures.

\begin{table}[htp]
\begin{tabular}{ |c|c|c|c| }
\hline
Material & Load & Young's modulus [GPa] & Poisson's ratio \\ \hline
NinjaFlex & Tensile & 0.1124 & 0.344 \\
 & Compressive & 0.15 & 0.340 \\ \hline
FlexPLA & Tensile & 1.743 & 0.342 \\
& Compressive & 1.4879 & - \\ \hline
Hard PLA  & Tensile & 7.23 & 0.0855 \\
 & Compressive & 3.1613 & - \\ \hline
\end{tabular}
\caption{\textbf{Base material properties.} Data was measured from solid printed samples of each material, under both tension and compression. }
\label{tab:base}
\end{table}

Prepared samples are put under tensile and compressive load by an Instron Electropuls E1000 testing machine, which applied a constant strain-rate of 0.02 mm/sec up to 1 kN force limitation. The amount of applied forced is recorded every 0.002 seconds and a gray scale camera (Dalsa Genie HM1024) records pictures every second.

\subsection*{Simulation}

The lattices are modelled as simple elastic networks \cite{feng1984percolation,mao2013effective} to capture the effects of the geometry on the Poisson's ratio. We simulate the response of 
a lattice under tension using molecular dynamics, with LAMMPS \cite{Plimpton1995}. A harmonic spring is placed between every vertex, with the energy of the spring described 
by $U = k_B(r-r_0)^2$, where $r$ is the bond length and $r_0$ is the equilibrium bond length. Additionally, a harmonic angular spring is placed between each of these bonds, described
by $U = k_Al(\theta-\theta_0)^2$, where $\theta$ is the bond angle and $\theta_0$ is the equilibrium bond angle. Tension was applied by fixing the velocity of the top row of vertices, 
while clamping the position of the bottom row of vertices. The fitting parameters were the bond ($k_B$) and angle ($k_A$) spring constants, however only their ratio affected the results, 
provided that the imposed tension velocity was sufficiently slow. The simulations are only performed in tension, to avoid any effects of contact between links.

To allow the links of the lattice to bend, necessary to fit the behavior of the experimental sample at high strains, each link is replaced by 4 equal length bonds in series. An angular spring is added at every vertex along the link, described by $U=k_Ll\Phi^2$, where the angle $\Phi$ is straight in equilibrium and the spring stiffness is $k_L$. The spring constants are fit to the experimental results, and represent the material properties. The energy of the full lattice is given by $U = k_B\sum_i(r_i-r_{0i})^2+k_Al\sum_j(\theta_j-\theta_{0j})^2+k_Ll\sum_m\Phi_{m}^2$ where the sums are taken over all bonds $i$, all angles $j$ and all link angles $m$. Comparing this constitutive equation to the definition of the Young's modulus, we can relate the spring constant used to the Young's modulus of a single link as $E=2k_Br_0/A_L$, where $A_L=0.6\text{mm}^2$ is the cross-sectional area of one link in the experimental sample. The bending modulus of our link with discrete nodes is given by $B = k_Lr_0l^2/2$, where $r_0$ is the equilibrium length of the full link.

The ratio of the bending and stretching modulus can be calculated for a simulated link as $B/E = k_Ll^2A_L/(4k_B)$. When the ratio of bond and link spring constants is fit to give the best match between simulation and experimental results ($k_L/k_B=0.0001$), we have $B/E=1.8\times10^{-15}\text{m}^2$, where $l=1$cm. This ratio can also be calculated directly from the area moment of inertia if we consider each link as a beam, $B/E = bh^3/12$. In our experimental samples, the thickness is $b=1.5$mm and the width is $h=0.4$mm, so $B/E = 8.0\times10^{-15}\text{m}^2$. This indicates the choice of spring constants to describe this system is reasonably consistent with the beam model.

\begin{figure}[ht]
\centering
\includegraphics[width = 10cm]{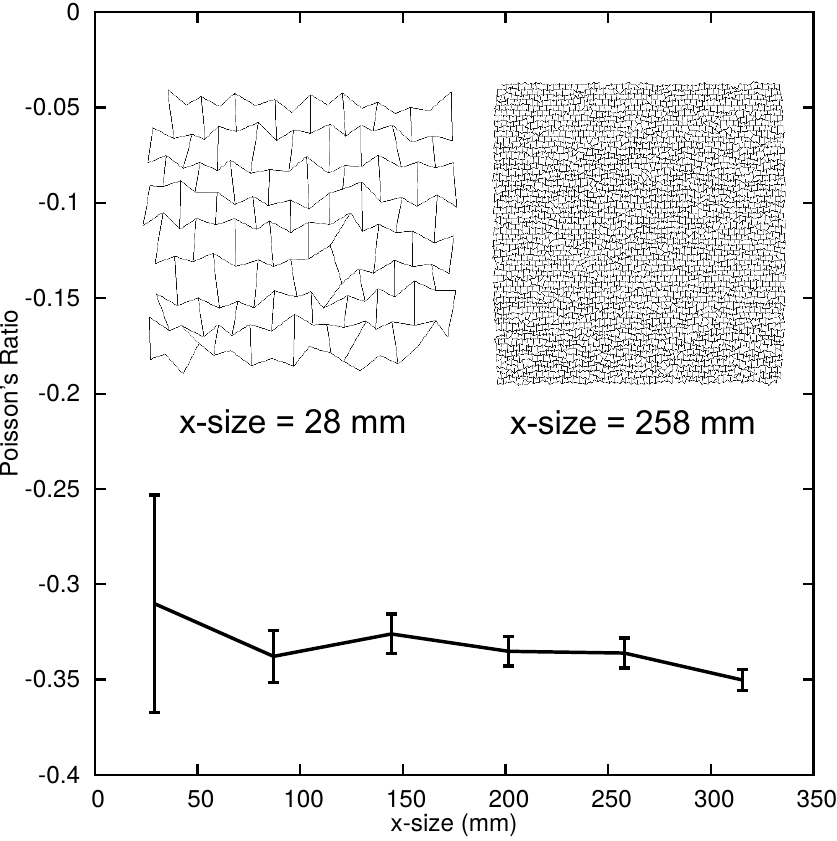}
\caption{\textbf{Simulated size dependence.} Poisson's ratio measured in simulations of different lattice sizes. Error bars are the standard deviation from 100 realizations of the disorder. All geometries were disordered with $d = 0.9$. Insert shows example geometries for two lattice sizes. \label{fig:sim}}

\end{figure}

The effect of the sample size on the Poisson's ratio is measured in simulations, and found to have no observable impact, Fig. \ref{fig:sim}.

\section*{\label{sec:results}Results}
\subsection*{\label{sec:elastic}Elastic properties}

\begin{figure}

\begin{centering}
\includegraphics[width=10cm]{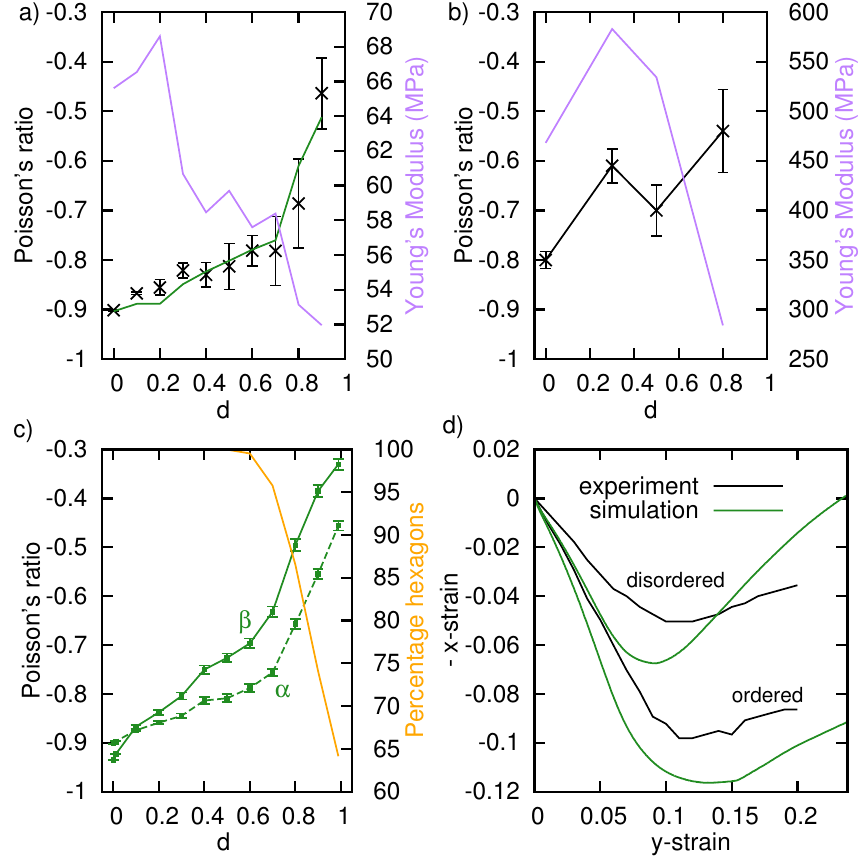}
\caption{\textbf{Behavior of flexible re-entrant lattices.} 
a) \textbf{Tension.} Experimental Poisson's ratio in black, with simulation values from the same geometries in green. Tension applied in the $\beta$ direction. Error bars are from  variation in the middle rows of the lattice. The experimental Young's modulus is in purple. b) \textbf{Compression.} Poisson's ratio for samples under compression in the $\beta$ direction in black, Young's modulus shown in purple. c)\textbf{ Large scale simulations of tension.} Poisson's ratio in green with error bars from 100 repetitions, the average percentage of hexagons in yellow. d) \textbf{Details of tension.} $y$-strain vs $-x$-strain for tension in the $\beta$ direction. In the disordered sample $d=0.9$.
\label{fig:poisson}}

\par\end{centering}
\end{figure}

We find that for all values of disorder, the Poisson's ratio remains negative under both tension, Fig. \ref{fig:poisson}(a), and compression, Fig. \ref{fig:poisson}(b). The lattice dimensions used are 7.5$\times$9$\times$0.15 cm and 4.5$\times$4.5$\times$1.0 cm respectively. The tension simulation uses the parameters $k_A/k_B=0.003$, and the same geometry as the experiment. Generally the Poisson's ratio increases with disorder (the structure becomes less auxetic). Under tension, for both the simulation and experimental results, this trend becomes more evident for values of d greater than 0.7. In both cases we also plot the Young's modulus, and find that it varies more with disorder under compression than tension. 

The trend in Poisson's ratio with disorder is further explored in simulations of larger lattice sizes (using a 16$\times$16 cell lattice rather than the 9$\times$9 cell experimental size), with 100 realizations of each value of disorder, plotted in Fig. \ref{fig:poisson}(c). There is no observable variation in Poisson's ratio with sample size (Fig. \ref{fig:sim}), and thus these results can be compared with the experiment, and we expect that the experimental results can be scaled up to larger lattice sizes. Also included in this plot is the percentage of hexagons, which classifies the number of topological defects. The point where the Poisson's ratio starts to change more rapidly with disorder corresponds to the point where the topology of the geometry first changes. Small changes in topology result in large changes in the Poisson's ratio. 

The Poisson's ratio is the slope of the $y$-strain vs. $-x$-strain plot, included as Fig. \ref{fig:poisson}(d) for the lattice under tension. The dimensions of this lattice are 7.5$\times$9$\times$0.15 cm. We see that for both the disordered and ordered samples, the lattice is only auxetic initially, before a turning point where the Poisson's ratio becomes positive. The disordered sample is auxetic over a smaller range of strains than the ordered structure. To fit this behavior after the auxetic regime in simulation, we had to allow each rib to bend in-plane. This was achieved by adding more nodes to the elastic network (see methods). The simulation parameters are  $k_A/k_B=0.005$ and $k_L/k_B=0.0001$.

\begin{figure}[ht]
\centering
\includegraphics[width=10cm]{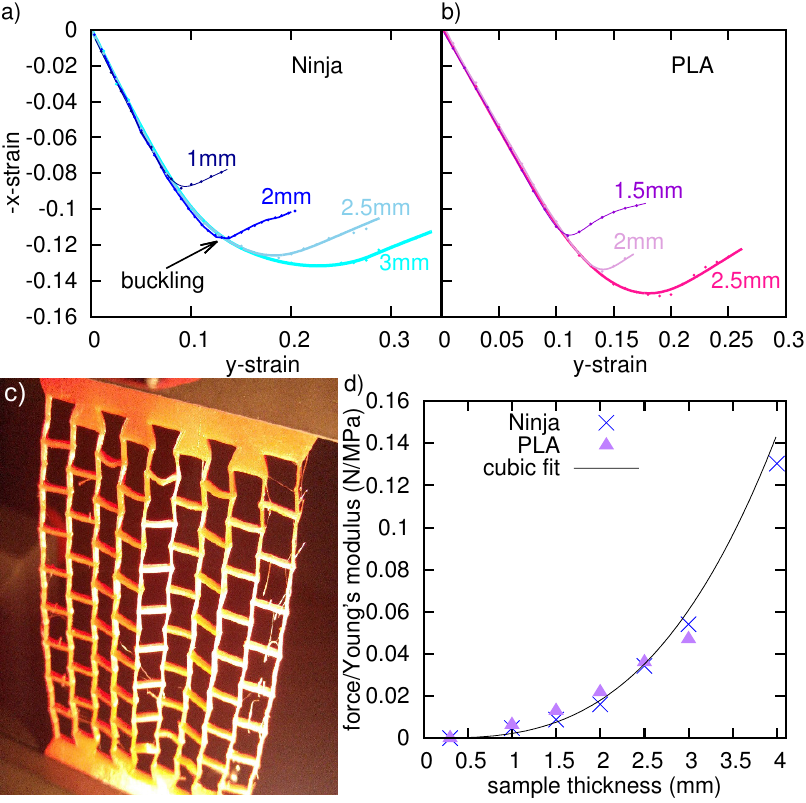}
\caption{\textbf{Out of plane buckling under tension.} a) Onset of buckling in the $y$-strain vs $-x$-strain plots, shown for NinjaFlex samples of different thicknesses. b) FlexPLA samples. c) Experimental sample buckles in a wave pattern, with alternating vertical strips in front of and behind the clamped plane. d) Sample thickness plotted against normalized force at the onset of buckling. A cubic fit is plotted in black. \label{fig:Buck}}
\end{figure}

The change to non-auxetic behavior during the tension experiments, plotted for two materials in Fig. \ref{fig:Buck}(a) and 3(b), corresponds to the sample buckling out of the plane, pictured in Fig. \ref{fig:Buck}(c). The sample size used in these experiments is 7.5$\times$10.5 cm. The point where the buckling starts, and so the range of strains over which the sample is auxetic, is highly dependent on the sample thickness. While buckling in systems under tension is rare \cite{zaccaria2011structures}, it has been observed when the boundaries are clamped \cite{cerda2002thin}, and its onset depends cubicly on sample thickness and linearly on the Young's modulus \cite{Friedl2000}. We observed the same relationship, plotted in Fig. \ref{fig:Buck}(d). While this behaviour may need to be controlled to measure the evolution of the Poisson's ratio, in this study all reported Poisson's ratios are measured before the onset of buckling. 

\subsection*{\label{sec:fracture}Lattice failure}

\begin{figure}[ht]
\centering
\includegraphics[scale = 1.0]{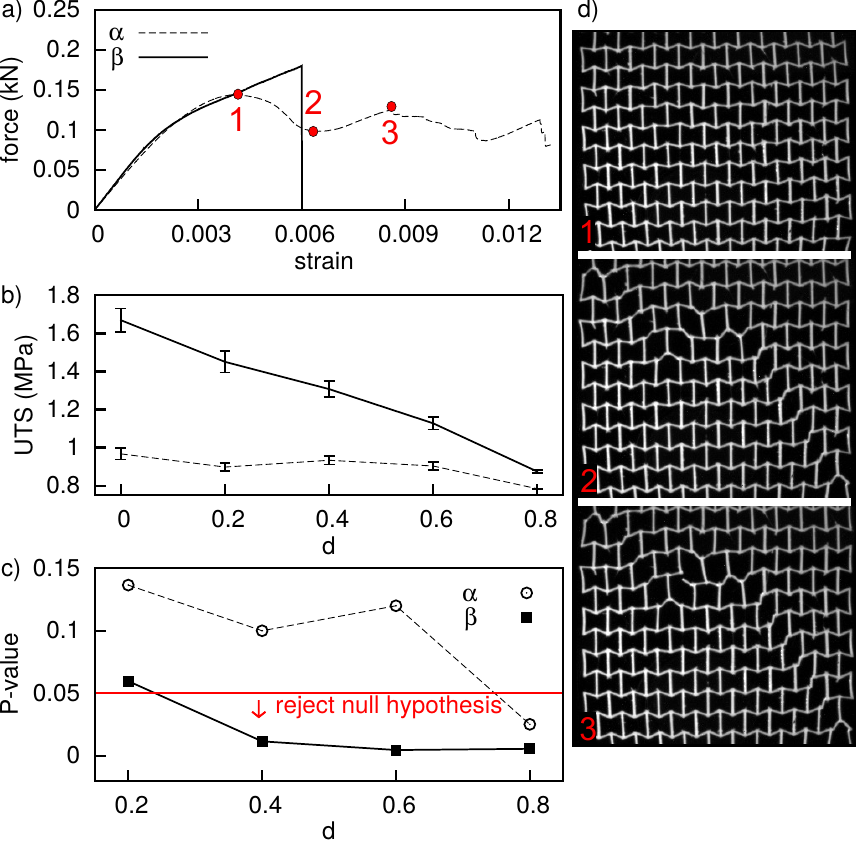}
\caption{\textbf{Brittle lattice under tension. }a) Force strain plot of ordered lattices. The numbered points on the $\alpha$ curve correspond to the pictures in d). b) The UTS is plotted as a function of disorder. Error bars are calculated from 5 repeats. c) Statistical analysis of UTS measurements. For each data point, a t-test was performed, with the null hypothesis that disorder has no effect on the UTS. d) Snapshots from the tension experiment on the $\alpha$ lattice.  \label{fig:fracTen}}

\end{figure}

\begin{figure}[ht]
\centering
\includegraphics[width=10cm]{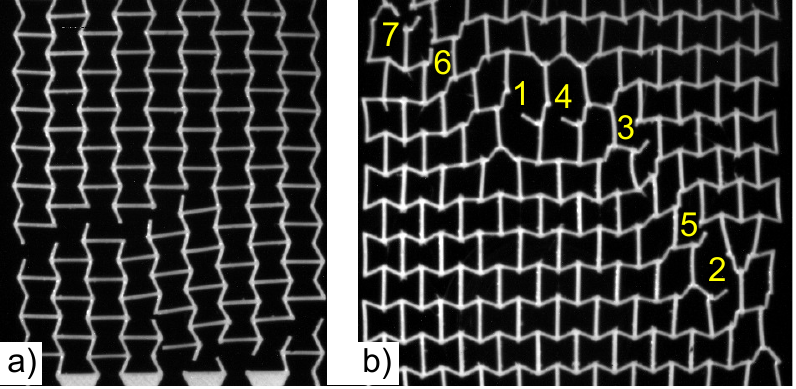}
\caption{ \textbf{Fracture of ordered lattices.} a) Photo of the final structure of an ordered experimental sample after tension has been applied in the $\beta$ direction. b) Final structure of an experimental sample after tension has been applied in the $\alpha$ direction. The broken bonds are numbered by the order in which they broke. \label{fig:fracDets}}

\end{figure}

We investigate the fracture properties of the auxetic lattices with samples printed in hard PLA (see methods), a more brittle filament. The mechanism for fracture under tension in the two orientations we have considered is vastly different. Figure \ref{fig:fracTen}(a) shows that the fracture in the $\beta$ direction is sudden and brittle, while in the $\alpha$ direction ductile behavior is observed, with the formation of a shear band preceding the fracture of the structure, pictured in Fig. \ref{fig:fracTen}(d). This can be seen clearly in the supplementary videos. Snapshots of the final fractured structure for both cases are presented in  Fig. \ref{fig:fracDets}. Details of the fracture of the $\alpha$ sample can be seen from the order of bond fracture. In the $\beta$ direction, the sample broke too rapidly for the details to be observed. The dimensions of these samples are 9.5$\times$10.6$\times$0.15 cm for the $\alpha$ orientation and  7.4$\times$11.3$\times$0.15 cm for the $\beta$ orientation. 
The ultimate tensile strength (UTS) for both orientations is plotted in Fig. \ref{fig:fracTen}(b). The small error bars (from 5 reprints of each geometry) indicate that the fracture is extremely consistent amongst the repeated samples, and the print quality used is sufficiently high. Performing a t-test on the data (presented in \ref{fig:fracTen}(c)), we may conclude that the UTS in the $\beta$ samples decreases with disorder for values of d as low as 0.4, where as in the $\alpha$ samples, no change is seen until $d=0.8$. This high value of disorder corresponds to the point where the geometry begins to contain topological defects. The start of the change in the UTS is perhaps expected given the respective failure characteristics, brittle and ductile.
\clearpage

\begin{figure}[ht]

\centering
\includegraphics[width=10cm]{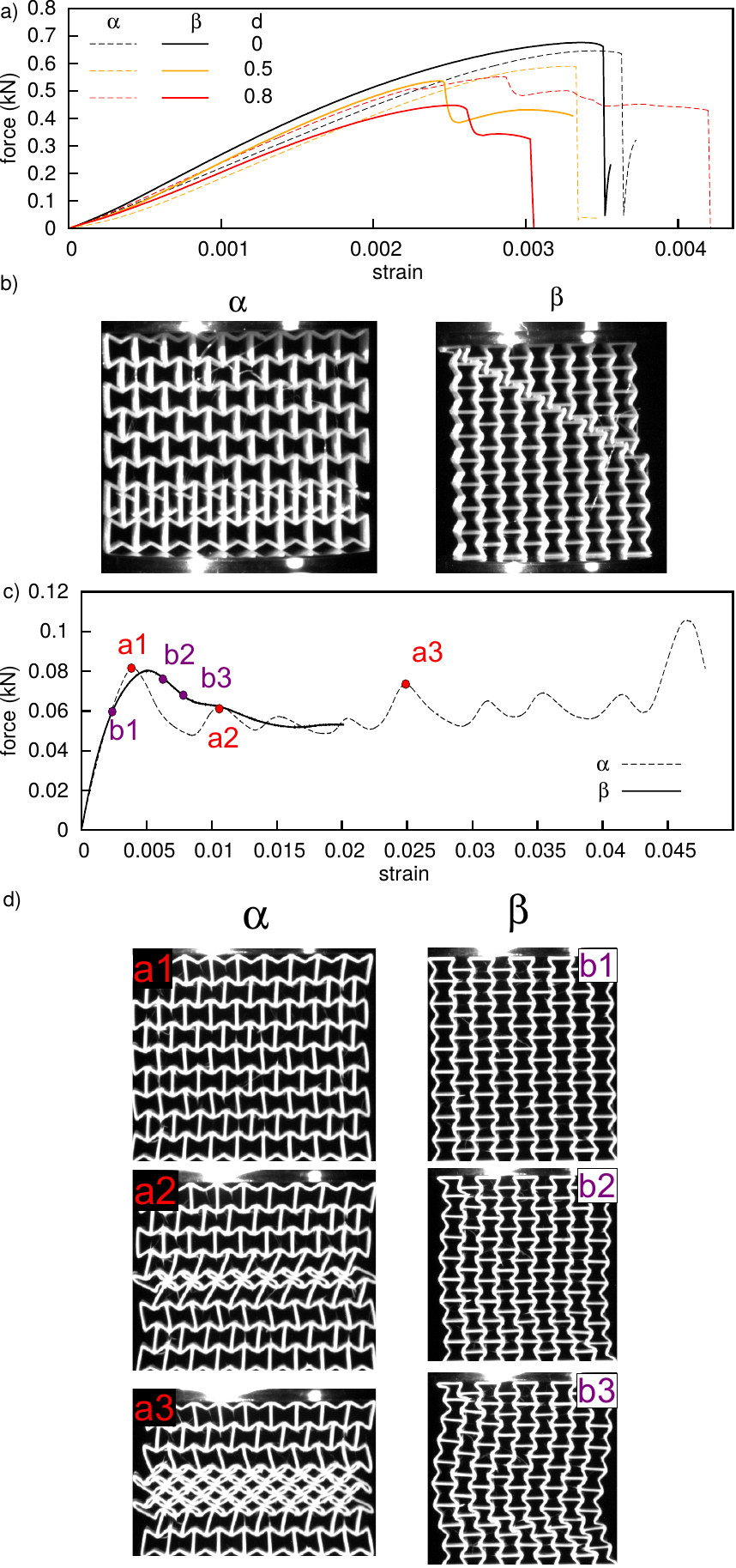}
\caption{ a) \textbf{Force-strain plots of brittle lattices under compression.} The $\alpha$ orientation is plotted as dashed lines, and the $\beta$ orientation is plotted in solid lines, for a range of values of d. b) \textbf{Photographs of fractured ordered lattices.} Both orientations are included. c) \textbf{Force-strain plots of flexible ordered lattice under compression.} Number points correspond to structures in d). d) \textbf{Photographs illustrating the compression process.} \label{fig:fracComp}}
\end{figure}

The force-strain plot of the brittle sample under compression are shown in Fig. \ref{fig:fracComp}(a).
They show little dependence on the orientation of the lattice or the amount of disorder. All samples exhibit brittle failure, though the disordered structures fail at slightly smaller strains. In all cases fracture occurs while the sample is in the auxetic regime. The fractured structures are pictured in Fig. \ref{fig:fracComp}(b).

In flexible samples, compression does not result in fracture. The force strain curves of flexible samples, shown in Fig. \ref{fig:fracComp}(b), illustrate the difference in the response to compression in each orientation. In the $\alpha$ direction each peak in the plot corresponds to one
row of the re-entrant lattice collapsing, exhibiting similar low shear resistance to the tensile experiment. 
In contrast, the force strain plot in the $\beta$ orientation is simpler, with the lattice first compressing symmetrically in the linear regime, before long wavelength in-plane buckling is observed. Snapshots of these processes are included in Fig. \ref{fig:fracComp}(d). For both material types in compression, the $\alpha$ sample has the dimensions 4.9$\times$4.4$\times$1.0 cm, and the $\beta$ sample is 4.4$\times$4.9$\times$1.0 cm.

\clearpage

\section*{\label{sec:conc}Summary and conclusions}

In this work we have thoroughly investigated the re-entrant hexagonal lattice structure in the presence of a controlled degree of disorder. We have measured its elastic behavior under both compression and tension, and where applicable we have confirmed the effects of the geometry with computer simulation. Unlike previous studies of disordered re-entrant hexagonal lattices \cite{Liu2014,Mukhopadhyay2016}, we find the addition of disorder does not have a large effect on the Poisson's ratio, unless the topology is modified. Disorder does not have a significant effect on the mechanism of fracture. However, the orientation of applied tension is important, where we see brittle failure in one direction but ductile failure in the other, and a notable difference in the sensitivity to disorder. 

This improved understanding of the elastic and fracture properties of disordered re-entrant geometries could contribute to the rational design of auxetic metamaterials. We expect that the three-dimensional case should exhibit similar robustness to the presence of disorder with a similar qualitative change when the lattice topology is finally altered. Though the strength properties of these systems are robust against the introduction of disorder - in contrast to usual lattice models of fracture and brittle materials \cite{Alava2006} - it would also be interesting to study in detail the relation of crack nucleation and growth to the topology of the local geometry. Further tuning or design using disorder might give rise to the possibility of the control of crack patterns and robustness.

\section*{Acknowledgments:}
 We thank Daniel Rayneau-Kirkhope (Aalto University) for useful comments.

\section*{Authors contributions}
MH performed experiments, CP performed numerical simulations, MJA and SZ coordinated and designed the project. All the authors wrote the paper.

\section*{Additional information}

\textbf{Competing financial interests} The authors declare no competing financial interests.


\end{document}